\documentclass[prb,twocolumn,superscriptaddress,showpacs,preprintnumbers,eqsecnum,amsmath,amssymb]{revtex4}

\begin{document}

\title{\bf Bell's Inequality Violation (BIQV) with Non-Negative Wigner Function}

\author{M. Revzen}
\affiliation{Department of Physics, Technion - Israel Institute of Technology, Haifa 32000, Israel}

\author{P.  A. Mello}
\affiliation{Instituto de F\'{\i}sica, Universidad Nacional Aut\'{o}noma de
M\'{e}xico, 01000 M\'{e}xico Distrito Federal, M\'{e}xico.}

\author{A. Mann}
\affiliation{Department of Physics, Technion - Israel Institute of Technology, Haifa 32000, Israel}

\author{ L. M. Johansen}
\affiliation{Department of Technology, Buskerud University College, Kongsberg, Norway}

\date{\today}

\begin{abstract}
A Bell inequality violation (BIQV) allowed by the two-mode
squeezed state
(TMSS), whose Wigner function is nonnegative, is shown to hold only for correlations
among dynamical variables (DV) that cannot be interpreted via
a local hidden variable (LHV) theory.

Explicit calculations and interpretation are given for Bell's suggestion that
the EPR (Einstein, Podolsky and Rosen) state will not allow for BIQV
in conjuction with its Wigner representative state being nonnegative.

It is argued that Bell's theorem disallowing the violation of Bell's inequality within a local hidden-variable
theory depends on the DV's having a definite value --assigned by the LHV-- even when they
cannot be simultaneously measured.
The analysis leads us to conclude that BIQV is to be associated with
endowing these definite values to the DV's and {\it not} with their
locality attributes.
\end{abstract}
\maketitle

\section{Introduction}
\label{intro}

In his article entitled (scented with an impish whiff)
``EPR (= Einstein, Podolsky and Rosen) correlations and EPW (=Eugene Paul Wigner) Distributions'',
Bell \cite{bell2}
studied the possibility of underpinning quantum theory with local hidden variables (LHV's) \cite{bell1}
in the case of two spinless particles.
He analyzed the correlations arising from measurements of positions of these
particles in free space --a situation closer to the original one envisaged by EPR \cite{rosen}--
utilizing the fact that Wigner's distribution \cite{wigner} simulates a local ``classical" model
of such correlations in phase space.
Bell suggested \cite{bell2} that the
nonnegativity of the Wigner function for certain quantum-mechanical states would preclude
Bell's inequality violation (BIQV)
with such states when one considers the correlations constructed from a dichotomous variable
defined as the sign of the coordinates of the particles.

We first recall a few properties of the Wigner function \cite{ulf}.
One can show that the expectation value of any operator ${\hat A}$
in a state defined by the density matrix ${\hat \rho}$ can be expressed as
\begin{equation}
{\rm Tr} ({\hat \rho}{\hat A}) = \int d\lambda W_{\hat{\rho}}(\lambda)W_{\hat{A}}(\lambda),
\label{wig}
\end{equation}
where $W_{\hat{Q}}(\lambda)$  is the Wigner representative of the quantal operator $\hat{Q}$
defined in Eq. (\ref{WQ}) below, and $\lambda$ designates the appropriate phase space coordinates, i.e.,
$\lambda = ({\bf q}, {\bf p}) = (q_1, \cdots, q_n, p_1, \cdots, p_n)$, $n$ being the number of degrees of freedom.
It should be noted that in Bell's considerations of LHV's, the values of the observables obey the
so-called Bell's factorization \cite{bell1,shimony2}, which leaves the value of each observable
independent of the ``setting" of the other.
In the expressions for two-particle correlations in terms of the Wigner representatives, 
when each of the DV's
depends on its own phase-space coordinates, this factorization is satisfied automatically.
This is our justification for referring to the description in terms of the Wigner function as {\it local}
\cite{bell2}.

We illustrate the above considerations using a two-mode squeezed state (TMSS) $|\zeta \rangle$,
defined as
\begin{equation}
|\zeta \rangle\;=\;\exp^{\zeta (a^{\dagger}b^{\dagger}\;-\;ab)}|00\rangle\;\equiv\;S(\zeta)|00\rangle;
\label{tmss}
\end{equation}
this equation defines the operator $S$.
Here,  the operators $a, \;a^{\dagger}$ refer to the beam of channel 1, while
$b,\;b^{\dagger}$ refer to those of the second channel \cite{faqir}.
In the limit of the squeezing parameter $\zeta$
increasing without limit, the state (\ref{tmss}) approaches the EPR state \cite{rosen}
$|EPR \rangle = \delta(q_{1}\;-\;q_{2})$ (here the subscripts refer to
the channels), as can be readily seen writing the state (\ref{tmss}) in the coordinate representation
as (we use well known normal ordering formula \cite{potasek})
\begin{eqnarray}
\langle q_{1}q_{2}|\zeta \rangle 
&=& \frac{1} {\cosh\zeta}\sum_{n=0}\;\tanh^{n}\zeta\;
\langle q_{1}q_{2}|nn\rangle\;
\nonumber \\
&{\buildrel \zeta \rightarrow\infty \over \longrightarrow}&
\sim \delta(q_{1}\;-\;q_{2}).
\end{eqnarray}
Now, the Wigner function, $
W_{\zeta}$, of the TMSS is given by \cite{wodkiewicz}
\begin{eqnarray}
&&W_{\zeta}(q_{1},q_{2},p_{1},p_{2})
\nonumber \\
&&\;\;\;\;\;\;\;=\frac{1}{\pi ^2}
\exp\left[-\cosh(2\zeta)\left(q_{1}^{2}+q_{2}^{2}+p_{1}^{2}+p_{2}^{2}\right) \right.
\nonumber \\
&&\;\;\;\;\;\;\;\;\;\; - 2\sinh(2\zeta)\left(q_{1}q_{2}-p_{1}p_{2}\right)\Big].
\label{W tmss}
\end{eqnarray}
and is clearly {\it nonnegative} for all $q$'s and $p$'s, and thus may be
considered as a distribution in phase space $(q_{1},q_{2},p_{1}, p_{2})$ associated with the state
$|\zeta \rangle$. Thus we may refer to the variables $(q_{1},q_{2},p_{1}, p_{2})$
as LHV's and correlations weighed with
$W_{\zeta}(q_{1},q_{2},p_{1}, p_{2})$ should preclude BIQV for dynamical
variables (DV's)  for which this may be a legitimate view \cite{70}.

As was mentioned above, Bell suggested \cite{bell2} that the
nonnegativity of the Wigner function of the EPR state would preclude BIQV
with this state when one considers the correlations of a dichotomous variable defined as
the sign of the coordinates of the particles.
The correlations considered in \cite{bell2}
are those that are involved in the CHSH \cite{shimony1} inequality, i.e.,
the inequality that is often studied in terms of the Bell operator
\cite{sam}.
(In the present paper, Bell's inequality and BIQV refer to this CHSH
inequality.)
Bell's original argument that nonnegativity of Wigner's function suffices
to preclude BIQV was shown to be inaccurate in \cite{lars}, where
difficulties in handling normalization of the EPR state considered by Bell  was shown to
involve a misleading factor.

The TMSS's were studied extensively since the early eighties in connection
with BIQV in general and, in particular, for their connection to the EPR state
\cite{reid,walls,grangier,mandel,ou1,ou2,santos}.
These studies focused
on the polarization as the observable (= dynamical variable, DV).
Banaszek and Wodkiewicz \cite{wodkiewicz} noted that {\it while
the Wigner function of the TMSS is non-negative, it allows for BIQV}, when the
dynamic variable involved in the correlations is the parity. Their study was
extended by Chen et al. \cite{zeng} who showed, by using appropriately
 defined spin-like DV's which together with the parity operator close an $SU(2)$
algebra, that the TMSS, $|\zeta \rangle$, allows the maximal possible
\cite{cirelson, landau} BIQV for $\zeta \rightarrow \infty$, i.e., when it is maximally
entangled \cite{entangled} and, as stated above, it tends to the EPR state.
An alternative parametrization (termed configurational)
to the spin-like operators was given in \cite{gour}.
This choice of DV is more convenient for our analysis as it involves the
DV considered by Bell and admits simple interpretation.

Our study aims at clarifying the relation between the non-negative
Wigner function of the TMSS, $|\zeta \rangle$, for all values of $\zeta$,
the DV's involved in the CHSH inequality \cite{shimony1,sam} and the
possibility of BIQV. The latter, by Bell's theorem \cite{bell1,shimony2},
prohibits the underpinning of the theory with a LHV theory. Note that
this attribute (non-negativity) of the Wigner function depends on variables 
over which it is defined \cite{agrawal}.

The paper is organized as follows.
In the next section we describe the properties that
should be required of a QM problem in order that its translation in terms of Wigner representatives
can be legitimately considered as a LHV theory.
We then divide the problem indicated in the last paragraph into three levels. 
The first
level, which the works hitherto were addressed to, is to consider BIQV with
the TMSS, viz., with a state with a non-negative Wigner function. 
In this connection we give, in Sec. \ref{eprw}, a brief review of Chen et al. \cite{zeng} considerations and
those of ref. \cite{gour}. We argue that the former approach \cite{zeng} involves,
exclusively, DV's whose Wigner representatives
are physically unsuitable for allowing a LHV theory underpinning
(in addition, they
do not fulfill the property of
boundedness, a mathematical condition that enters the derivation of BIQ).
Such DV's that are ineligible for a LHV theory in phase space
(the domain of Wigner's function \cite{agrawal}) are termed {\it improper} or {\it dispersive} DV's - the
definition of these terms and their justification is also included in Sec. \ref{HVWT}.
We then consider the next level of the problem, viz., where
in addition to having the non-negative Wigner function of $|\zeta \rangle$, we
have a DV that is proper (or nondispersive), i.e., one that can be accounted for by
the LHV that the phase space provides (indeed it is the very one
considered by Bell \cite{bell2}: the sign of the coordinate of the particle).
However, we show that its mates, i.e., its rotated (we use here the spin analogy) partner(s) which, with it, must be
present in the Bell operator \cite{sam}, are dispersive
(they are also not bounded) and hence, again, no LHV theory can be
sustained here. 
We also discuss the alternative approach of retaining the
original DV and rotating the wave function and show that in this case it leads
to a {\it non} non-negative Wigner function. In section \ref{bilinear}  we finally
study the last level which is the one considered by Bell.
In addition to having
the non-negative Wigner function and the proper DV - its ``rotated'' mates
are now gotten by time evolution with a ``free'' Hamiltonian. For this case
we show that the evolved DV remains non-dispersive, or
alternatively (perhaps less surprising), the ``rotated'' wave function continues to give rise to a
non-negative Wigner function.
 We thus arrive at the conclusion that Bell's expectation \cite{bell2} that the EPR state will not allow BIQV
is confirmed.
However, our approach underscores the importance of the perhaps
not sufficiently stressed
assumption involved in the derivation of Bell's inequalities
\cite{bell1, shimony1},
viz that the LHV theory be such that the DV are defined
simultaneousely even when they cannot be measured simultaneousely.
This point was noted before \cite{wigajp,sakurai,willy,milos,stapp,laloe}. Indeed, 
 such a requirement tantamounts to having the 
LHV endowing physical reality (in the EPR sense \cite{rosen}) to the DV's
measureable attributes.

To remain close to the formalism as discussed by Bell \cite{bell2} we shall
throughout refer to variation of the DV's as ``evolution".
This retains complete generality, since to define the evolution we can choose
a Hamiltonian leading to the required variation.

\section{Hidden Variables and Wigner's transform}
\label{HVWT}

We consider bounded QM operators $\hat{A}$ associated with DV's 
for a given physical system, with
eigenvalues $a_{n}$.  By a proper rescaling, we can always have
\begin{equation}
|a_n|\leq 1 .
\label{bounded e-values}
\end{equation}
In a HV theory we assume that we have variables $\lambda $ endowed with a probability distribution
\begin{equation}
\rho (\lambda )\geq 0 \; ,
\label{rho>0}
\end{equation}
such that to every operator $\hat{A}$ we associate, according to some recipe,
a function $A(\lambda )$ --a ``representative" of the DV in terms of the hidden variable $\lambda $--
that takes on, as its possible values, the eigenvalues $a_n$.
When this is feasible, we say that we are dealing with a ``proper" dynamical variable (PDV).
Notice that this property implies that if $A(\lambda )$ is the representative of the operator $\hat{A}$,
then $A^k(\lambda )$ is the representative of the operator $\hat{A}^k$, where $k$ is an integer.
We then speak of a ``non-dispersive" DV.
As a consequence, the $A(\lambda )$'s are bounded as
\begin{equation}
|A(\lambda )| \leq 1 .
\label{A bounded}
\end{equation}
In a two-particle problem, if the DV $\hat{A}$ is associated with particle 1 and $\hat{B}$ with particle 2,
the requirement that $A(\lambda )$ be independent of the setting ${\bf b}$ of the instrument that measures particle 2
and $B(\lambda )$ be independent of the setting ${\bf a}$ of the instrument that measures particle 1 makes the theory
a LHV theory.
For this two-particle problem we now introduce two other DV's, $\hat{A}'$ and $\hat{B}'$,
associated with particles 1 and 2, respectively, and
not commuting, in general, with $\hat{A}$ and $\hat{B}$, respectively.
To these new DV's we associate the functions $A'(\lambda )$ and $B'(\lambda )$, respectively.
Notice that the functions $A(\lambda )$ and $A'(\lambda )$ for particle 1 
(and similarly $B'(\lambda )$ and $B'(\lambda )$ for particle 2)
{\it assign a definite value to the two DV's,
whether they can be measured simultaneously or not}.
Then one can prove the CHSH inequality
\begin{equation}
|\left\langle {\cal B}(\lambda )  \right\rangle|
\equiv \left| \int  {\cal B}(\lambda ) \rho (\lambda ) d \lambda \right|  \leq 2 ,
\label{chsh}
\end{equation}
where  ${\cal B}$ is given by
\begin{equation}
{\cal B} = A(\lambda )B(\lambda ) + A(\lambda )B'(\lambda ) +A'(\lambda )B(\lambda ) - A'(\lambda )B'(\lambda ) .
\label{Bell fctn}
\end{equation}
As we mentioned in the Introduction, we shall call the above inequality BIQ.
In other words, dealing with PDV's implies (\ref{A bounded}) which, in turn, implies BIQ:
\begin{equation}
PDV \Rightarrow (\ref{A bounded}) \Rightarrow BIQ  ,
\label{PDV-B-BIQ}
\end{equation}
so that
\begin{equation}
PDV \Rightarrow  BIQ  .
\label{PDV-BIQ}
\end{equation}
Conversely, in a HV model in which (\ref{rho>0}) is fulfilled, a violation of BIQ (to be called BIQV) implies that (\ref{A bounded}) is not
fulfilled, and hence that we are not dealing with PDV's, i.e.
\begin{equation}
BIQV\Rightarrow \overline{(\ref{A bounded})} \Rightarrow   \overline{PDV},
\label{BIQV-B-PDV}
\end{equation}
so that
\begin{equation}
BIQV\Rightarrow \overline{PDV} .
\label{BIQV-PDV}
\end{equation}
(The bar on a proposition indicates its negation.)
We mentioned these conditions with some care because of the various applications that we shall be concerned
with in the following sections.

Let us mention that when we deal with dichotomous variables, i.e., with operators having only two eigenvalues ($\pm 1$),
one can prove that the QM expectation value for any two-particle state $\left| \Psi\right\rangle$ of the Bell operator
\cite{sam}
\begin{equation}
\hat{{\cal B}} = \hat{A}\hat{B}  + \hat{A}\hat{B}' + \hat{A}'\hat{B} - \hat{A}'\hat{B}'
\label{Bell op}
\end{equation}
satisfies the Cirel'son inequality \cite{cirelson}
\begin{equation}
\left| \left\langle \Psi \right| \hat{{\cal B}} \left| \Psi\right\rangle  \right| \leq 2\sqrt{2}.
\label{cirelson}
\end{equation}

We now discuss a specific way of implementing the above LHV program in terms of the theory of
Wigner's transforms.
We define the Wigner representative $W_{\hat{Q}}(q,p)$ of the quantal operator $\hat{Q}$ (for one degree of freedom) as \cite{schleich1}
\begin{equation}
W_{\hat{Q}}(q, p)
=\int e^{-ip\cdot y}
\left\langle q + \frac 12   y \right| \hat{Q} \left|  q - \frac 12  y  \right\rangle dy,
\label{WQ}
\end{equation}
while the Wigner function for the density operator is defined with an extra factor of ${1\over 2\pi} $ for each degree of freedom, i.e for one degree of freedom:
\begin{equation}
W_{\hat{\rho}}(q, p)
={1 \over 2\pi}\int e^{-ip\cdot y}
\left\langle q + \frac 12   y \right| \hat{\rho} \left|  q - \frac 12  y  \right\rangle dy.
\label{Wden}
\end{equation}
Then one can prove that the expectation value of an operator $\hat{A}$ with the density matrix
$\hat{\rho }$ is \cite{schleich1}
\begin{equation}
{\rm Tr} (\hat{\rho}\hat{A}) =  \int  W_{\hat{\rho}}( q, p)W_{\hat{A}}( q, p)
dq dp.
\label{wig 1}
\end{equation}
One can easily see that $W_{\hat{Q}}( q, p)$ of Eq. (\ref{WQ}) can also be expressed as
\begin{subequations}
\begin{eqnarray}
W_{\hat{Q}}( q, p) &=& {\rm Tr}\left[  \hat{Q} \; \hat{\Omega } ( q, p) \right] ,
\label{WQ 1}
\\
\hat{\Omega}(q, p) &=& \int \left| q - \frac 12 y \right\rangle e^{-i p\cdot y} \left\langle q + \frac 12   y \right|dy  ,
\nonumber \\
\label{Omega}
\end{eqnarray}
\label{WQ 1'}
\end{subequations}
an expression that will be useful later.

It can be shown \cite{schleich} that the only wave function whose Wigner transform is non-negative
is a Gaussian:  in this case, the associated Wigner transform is apparently interpretable as a probability density
in phase space (see Eq. (\ref{rho>0})).
The TMSS of Eq. (\ref{tmss}) is an example where this interpretation is indeed feasible.
If, in addition, the Wigner representatives of the DV's under study are of the proper, or non-dispersive, 
nature required above,
we have a candidate for a LHV theory, where the LHV's are represented by the canonical variables $q$ and $p$.
It seems clear from the outset that it will be rather exceptional for a DV to fall into this category.
It is the purpose of the discussion that follows in the present section to identify a class of operators $\hat{A}$ that do correspond
to proper DV's. Although the analysis is certainly not exhaustive, it serves the purpose of indicating 
a number of sufficient conditions leading to PDV's.
For simplicity, the analysis will be restricted to systems with only one degree of freedom.

Consider a function $f(x)$, where $-\infty \leq x \leq \infty$ and the function is bounded as $|f(x)| \leq 1$.

1. We define the operator $\hat{A}_1=f(\hat{q})$ through its spectral representation as
\begin{equation}
\hat{A}_1=f(\hat{q})
=\int_{-\infty}^{\infty}\left|  q'   \right\rangle   f(q')    \left\langle q'  \right|  dq'   .
\label{A=f(q)}
\end{equation}
The eigenvalues of this operator are $f(x)$, so that its spectrum lies in the interval $[-1,1]$. 
For instance:

(a) $f(x)=\tanh x$ gives a continuous spectrum in the interval $[-1,1]$.

(b) $f(x)={\rm sgn} x$ (where the ${\rm sgn}$ function takes on the value $1$ for $x>0$ and $-1$ for $x<0$)
has a discrete spectrum, consisting of the two values $1$ and $-1$.

One can easily show that the Wigner transform of the operator $f(\hat{q})$ of Eq. (\ref{A=f(q)}) is
\begin{equation}
W_{f(\hat{q})}(q', p') = f(q'),
\label{Wf(q)}
\end{equation}
a function which takes on, as its values, precisely the eigenvalues of the operator $f(\hat{q})$.
According to our nomenclature, we are thus dealing with a PDV. In these examples we see
the non-dispersive property explicitly, since
\begin{equation}
W_{\left[f(\hat{q})\right]^k}(q', p')=\left[W_{f(\hat{q})}(q', p')\right]^k.
\end{equation}

2. Similar considerations apply to the operator $\hat{A}_2=f(\hat{p})$.

3. Another case, which is very relevant for our future considerations, is that of the operator
\begin{equation}
\hat{A}_3=f(\widehat{\bar{q}}) ,
\label{A=f(q')}
\end{equation}
where
\begin{equation}
\hat{\bar{q}} = a \hat{q} + b \hat{p},
\label{q'}
\end{equation}
($a$ and $b$ being numerical constants) is a linear combination of the position and
momentum operators $\hat{q}$ and $\hat{p}$.
If we add, to Eq. (\ref{q'}), the following one:
\begin{equation}
\hat{\bar{p}} = c \hat{q} + d \hat{p},
\label{p'}
\end{equation}
$c$ and $d$ being numerical constants satisfying the condition
\begin{equation}
ad-bc = 1,
\label{D=1}
\end{equation}
then the pair of equations (\ref{q'}) and (\ref{p'}) can be considered as a transformation from the
canonical position and momentum operators $\hat{q}$ and $\hat{p}$ to the new ones $\hat{\bar{q}}$ and
$\hat{\bar{p}}$.
Thanks to the condition (\ref{D=1}), the commutator $[\hat{q},\hat{p}]=[\hat{\bar{q}},\hat{\bar{p}}]=i$ is preserved and the transformation is
canonical: it is the quantum-mechanical counterpart 
\cite{moshinsky}
of the classical linear canonical transformation obtained from Eqs. (\ref{q'}) and (\ref{p'}) by removing
the ``hats" and considering the $q$, $p$, $\bar{q}$ and $\bar{p}$ as $c$-number canonical variables; in the classical
problem it is the Poisson bracket that is preserved by the transformation.

The operators $\hat{\bar{q}}$, $\hat{q}$ have the same spectrum,  and so do the operators
$\hat{\bar{p}}$, $\hat{p}$; we can thus relate the two members of each 
pair through the unitary transformation
\begin{subequations}
\begin{eqnarray}
\hat{\bar{q}} = V^{\dagger} \hat{q} V \\
\hat{\bar{p}} = V^{\dagger} \hat{p} V.
\label{q'=VqV}
\end{eqnarray}
\end{subequations}
The eigenstates of $\hat{\bar{q}}$ and $\hat{\bar{p}}$, to be designated by $\left| q' \right)$ and $\left| p'\right)$,
respectively, i.e.,
\begin{subequations}
\begin{eqnarray}
\hat{\bar{q}}\left| q' \right) &=& q' \left| q' \right) 
\label{ev eqn for q'}
\\
\hat{\bar{p}}\left| p' \right) &=& p' \left| p' \right) ,
\label{ev eqn for p'}
\end{eqnarray}
\end{subequations}
are related to the eigenstates $\left| q' \right\rangle$,  $\left| p' \right\rangle$ of $\hat{q}$ and $\hat{p}$,
respectively, as
\begin{subequations}
\begin{eqnarray}
\left| q' \right) &=& V^{\dagger}\left| q' \right\rangle
\label{e states of q vs q'}
\\
\left| p' \right) &=& V^{\dagger}\left| p' \right\rangle .
\label{e states of p vs '}
\end{eqnarray}
\end{subequations}

In terms of the eigenstates $\left| q' \right)$ of $\hat{\bar{q}}$, Eq. (\ref{ev eqn for q'}),
we can write the spectral representation of the operator
$\hat{A}_3$ of Eq. (\ref{A=f(q')}) as
\begin{equation}
\hat{A}_3=f(\hat{\bar{q}})
=\int_{-\infty}^{\infty}\left|  q'   \right)  f(q')    \left( q'  \right|  dq'  .
\label{spectral repr of A3}
\end{equation}
Using Eqs. (\ref{e states of q vs q'}) and (\ref{A=f(q)}), we can write further
\begin{subequations}
\begin{eqnarray}
\hat{A}_3
&=&f(\hat{\bar{q}})
= V^{\dagger}
\int_{-\infty}^{\infty}\left|  q'   \right\rangle  f(q')    \left\langle q'  \right|  V dq'
\\
&=& V^{\dagger} f(\hat{q}) V  .
\label{A3 vs A1}
\end{eqnarray}
\end{subequations}
From Eq. (\ref{spectral repr of A3}) we read off the eigenvalues of the operator $\hat{A}_3=f(\hat{\bar{q}})$
as $f(x)$, just as for  $\hat{A}_1=f(\hat{q})$: in point of fact, a unitary transformation (Eq. (\ref{A3 vs A1})
does not change the spectrum !

The next step is analyze the properties the Wigner transform of the operator $\hat{A}_3=f(\hat{\bar{q}})$.
We first make a more general statement: from Eqs. (\ref{WQ 1'}) one can show 
that the Wigner transform of two operators $\hat{A}$
and $V^{\dagger}\hat{A}V$, $V$ being the unitary operator discussed above, are related by
\begin{equation}
W_{V^{\dagger}\hat{A}V}(q', p') = W_{\hat{A}}(aq'+bp', cq' +dp').
\label{W(A") vs W(A)}
\end{equation}
In other words, if the operator $\hat{A}$ undergoes the unitary transformation $\hat{A} \Rightarrow V^{\dagger}\hat{A}V$,
the Wigner transform is affected precisely by the classical linear canonical transformation
of which Eq. (\ref{q'}) is the quantum-mechanical counterpart.
Now, if we apply this result to the operators $f(\hat{q})$ and $V^{\dagger} f(\hat{q}) V = f(a\hat{q} + b\hat{p})$
of Eq. (\ref{A3 vs A1}), we find
\begin{equation}
W_{f(a\hat{q} + b\hat{p})}(q', p') =W_{f(\hat{q})}(aq'+bp', cq' +dp'),
\label{W(f) vs W(f')}
\end{equation}
and, using Eq. (\ref{Wf(q)}) for the right-hand side, we finally obtain
\begin{equation}
W_{f(a\hat{q} + b\hat{p})}(q', p') = f(aq'+bp'),
\label{W(f) vs W(f') 1}
\end{equation}
which clearly reduces to Eq. (\ref{Wf(q)}) when $a=1$ and $b=0$.

Right after Eq. (\ref{A3 vs A1}) we identified the spectrum of $f(a\hat{q} + b\hat{p})$ as $f(x)$. Now, Eq. (\ref{W(f) vs W(f') 1})
tells us that the Wigner transform of this operator takes on, as its values, exactly the eigenvalues of the
quantum-mechanical operator: we are thus dealing with a PDV.
As a result, {\it we have found a class of observables}, i.e., $f(a\hat{q} + b\hat{p})$ which, together with their
Wigner transforms, i.e., $f(aq' + bp')$,
{\it may be termed PDV's}.

As an application,
suppose that we have a two-particle problem, with the Wigner distribution associated with the wave function
being non-negative. 
Suppose also that we choose, as the operators $\hat{A}$, $\hat{A}'$ to be associated with particle 1,
any two (in general non-commuting) of the proper ($\Rightarrow$ non-dispersive \cite{lars2}) DV's discussed above,
like $\hat{A}_1$, $\hat{A}_2$, or $\hat{A}_3$, and similarly for the operators $\hat{B}$, $\hat{B}'$
to be associated with particle 2.
Then the CHSH inequality (\ref{chsh}) must be fulfilled, according to the discussion given right before that equation.
In the presentation carried out in Sec. \ref{bilinear} below, $\hat{A}$ is taken as ${\rm sgn}(\hat{q})$, i.e., as
$\hat{A}_1$ above, Eqn. (\ref{A=f(q)}), case (b); $\hat{A}'$ is taken as $\hat{A}_3$ above, Eqn. (\ref{A=f(q')}),
again with $f(x)={\rm sgn}(x)$, for two options for the coefficients $a$ and $b$. Similar choices are made for
$\hat{B}$ and $\hat{B}'$.
For these cases, the validity of the CHSH inequality (\ref{chsh}) is verified explicitly
in Sec. \ref{bilinear}.

In contrast, it is easy to give examples of DV's that do not fulfill the above property of having a Wigner function taking,
as its values, the eigenvalues of the quantum operator.
For instance, for the observable
\begin{equation}
\hat{A} = \frac 12 \left( \hat{p} ^2  +  \hat{q} ^2\right),
\label{A as ho}
\end{equation}
the quantum-mechanical spectrum is $n+1/2$ ($n=0,1,2,\cdots$).
(This spectrum is not bounded in the sense of (\ref{bounded e-values}); it just serves as an example to illustrate the point.)
In contrast, its Wigner transform is
\begin{equation}
W_{\frac 12 \left( \hat{p} ^2  +  \hat{q} ^2\right)}(q', p') = \frac 12 \left[ (p') ^2  +  (q') ^2\right],
\label{W for A as ho}
\end{equation}
which takes on {\it any} value in $\left[ 0, \infty \right]$:
the DV (\ref{A as ho}), together with its Wigner representative (\ref{W for A as ho}), is thus improper.
Some of the observables considered in Sec. \ref{eprw} below will, indeed, fail to be proper.

The idea of the present section has been to gain a panoramic view of the various particular cases that will be considered
in the rest of this paper.

Before turning to a study of these individual situations,
we mention in passing one further application of Eq. (\ref{W(A") vs W(A)}).
Consider the variation of ${\rm Tr}(\hat{\rho}\hat{A})$, Eq. (\ref{wig 1}), when
the operator $\hat{A}$ is subjected to the unitary transformation $\hat{A} \Rightarrow V^{\dagger}\hat{A}V$;
obviously, the same answer is obtained if, instead,
$\hat{\rho }$ is transformed as $\hat{\rho} \Rightarrow V \hat{\rho}V^{\dagger}$.
We can calculate the change of the Wigner representative of $\hat{\rho}$ from Eq. (\ref{W(A") vs W(A)}),
valid for any Hermitean operator, replacing $\hat{A}$ by $\hat{\rho }$ and $V$ by its inverse, with the result
\begin{equation}
W_{V\hat{\rho }V^{\dagger}}(q', p') = W_{\hat{\rho }}(dq'-bp', -cq' +ap')  ,
\label{W(rho') vs W(rho)}
\end{equation}
which will be useful later.

\section{The EPR-EPW Problem}
\label{eprw}

As outlined in the Introduction, we consider the so-called EPR-EPW problem \cite{bell2,lars}
in successive levels.
The first level is: Given a state, $|\zeta \rangle$ in our case, whose
Wigner representative function is non-negative, does such a state allow BIQV?

The answer to this was shown \cite{wodkiewicz,zeng} to be in the affirmative.
The DV considered was the parity, $S_{z}$  ($\hat{N}$ being the number
operator),
\begin{equation}
S_{z}\;\equiv\;\sum_{n=0}^{\infty}\Big[|2n+1\rangle\langle 2n+1|\;-\;
|2n \rangle \langle 2n|  \Big]\;=\;-(-1)^{\hat{N}}.
\end{equation}
In \cite{zeng}, ``rotated" parity operators were introduced:
\begin{equation}
S_{x}=\sum_{n=0}^{\infty}\Big[|2n+1 \rangle \langle 2n|\;+\;|2n \rangle
\langle 2n+1|\Big],
\end{equation}
\begin{equation}
S_{y}=i\sum_{n=0}^{\infty}\Big[ |2n \rangle \langle 2n+1|\;-\;|2n+1\rangle
\langle 2n|\Big].
\end{equation}
These operators close an $su(2)$ algebra and are viewed as 3-dimensional vector
operators. We may thus consider a ``rotation'' in parity space by, e.g.,
\begin{equation}
S_{x}'(\vartheta)
\;=\;e^{{i\vartheta \over 2} S_{z}}S_{x}e^{{-i\vartheta \over 2}S_{z}}\;=\;
S_{x}\cos\vartheta - S_{y}\sin\vartheta
\;=\;
{\bf S\cdotp n}
\label{S'}
\end{equation}
with ${\bf n}$ a unit vector which, in this case, is in the ``$x-y$" plane of the parity
space. It will be convenient for us later to refer to the above as the
``time evolution" of $S_{x}$ under the ``Hamiltonian'' $S_{z}$ in Eq. (\ref{S'}): in this
way we refer to the ``rotation'' angle, $\vartheta$, as the time, $t$.
Sticking to the geometric notation, the Bell operator \cite{sam} is
(the superscript refer to the channels, $a, a^{\dagger}$ being channel 1 and
$b, b^{\dagger}$ channel 2)
\begin{eqnarray}
\hat{{\cal B}}
&=&{\bf S^{1}\cdotp n}\;{\bf S^{2}\cdotp m}
\;+\;{\bf S^{1}\cdotp n'}\;{\bf S^{2}\cdotp m}
\nonumber \\
&&\;\;\;\;\;+{\bf S^{1}\cdotp n}\;{\bf S^{2}\cdotp m'}
\;-\;{\bf S^{1}\cdotp n'}\;{\bf S^{2}\cdotp m'},
\end{eqnarray}
and the Bell inequality we study is
\begin{equation}
|\langle\hat{{\cal B}}\rangle|\;\leq\;2.
\end{equation}
Varying ${\bf n, n'}$ and ${\bf m, m'}$ to maximize $|\langle\hat{{\cal B}}\rangle|$
for the state $|\zeta \rangle$ we get
\cite{gour}
\begin{equation}
|\langle \zeta|\hat{{\cal B}}|\zeta \rangle|\;=\;2\sqrt{1 + F^{2}(\zeta)};
\end{equation}
\begin{equation}
F(\zeta)\;=\;\langle \zeta|\;S_{x}^{1}\;S_{x}^{2}\;|\zeta\rangle\;=\;
\tanh2\zeta.
\end{equation}
Thus the state $|\zeta \rangle$ allows BIQV, even though the Wigner function of
the corresponding density operator may be viewed as a probability density of
LHV (the phase space coordinates).
However, as was stressed in the
Introduction, this does not violate Bell's theorem which prohibits BIQV for
a LHV theory. Thus the correlations appearing in the Bell operator have the structure \cite{ulf}
\begin{eqnarray}
\langle \zeta|S_{z}^{1}S_{z}^{2}|\zeta \rangle
&=&\int_{-\infty}^{\infty}dp_{1}dq_{1}dp_{2}dq_{2}W_{\zeta}(p_{1},q_{1},p_{2},q_{2})
\nonumber \\
&&\hspace{1cm} \cdot W_{S_{z}^1}(p_{1},q_{1})  W_{S_{z}^2}(p_{2},q_{2})   .
\label{<SzSz>}
\end{eqnarray}
Here, the factorization of the Wigner function of the two channels is automatic.
As explained in detail in Sec. \ref{HVWT},
for the right-hand side of Eq. (\ref{<SzSz>}) to be interpretable as a LHV theory, aside from a
nonnegative Wigner function for the state, $W_{\zeta}$, we require that the
Wigner representative of the DV's, the $S_{z}$'s in this case,
give the observable values of these DV's, viz., the eigenvalues of the quantal
parity operator (for the phase point: $q$, $p$).
As already indicated, we refer to a DV with this property as a {\it proper} 
or {\it non}dispersive DV \cite{lars2}.
This is not the case for any of the parity operators, $S_{i}$ ($i=x,y,z$);
in fact, e.g., we can easily verify that
\begin{equation}
W_{S_{z}}(q,p)\;=\;-\pi\delta(\alpha)\;=\;-\pi\delta(q)\delta(p),\;\;\alpha\;=\;q\ +ip.
\label{W(Sz)}
\end{equation}
This clearly is not an eigenvalue of the parity operator (which is $\pm1$).
Thus in this case this DV is {\it im}proper or dispersive \cite{lars2}.
Therefore, we are not dealing here with a LHV theory.
(In addition, Eq. (\ref{W(Sz)}) makes it clear the assertion made in the Introduction
that the Wigner representative of $\hat{S}_z$ violates the property of boundedness.)

We have thus completed the discussion of the first level of the EPR-EPW
problem: nothing new was gained but we considered examples that will serve us
below.

The second level of the EPR-EPW problem is when, in addition to
having a nonnegative Wigner function for the state, we have a DV whose Wigner
representative is the value of the DV - i.e. it is a proper or nondipersive DV.
Would this situation allow BIQV? Would it conform to Bell's theorem?
Recently \cite{gour,70} an alternative configuration was discussed for the
parity operators. In this alternative configuration the operators are given in
the $q$ representation. Denoting the operators in this configuration by
$\Pi_{i} \;(i=x,y,z),$ we have,
\begin{equation}
\Pi_{z}\; \equiv\; -\int_{0}^{\infty}dq\Big[ |{\cal E}\rangle \langle {\cal E}|\;
-\;|{\cal O}\rangle\langle{\cal O}|\Big ] \;\;=\;\;S_{z};
\end{equation}
here,
\begin{equation}
|{\cal E}\rangle
\;=\;{1 \over \sqrt2}\Big[ |q\rangle+|-q\rangle\Big],\;\;\;
|{\cal O}\rangle\;=\;{1 \over \sqrt2}\Big[ |q\rangle-|-q\rangle\Big] ,
\end{equation}
so that
\begin{equation}
\Pi_{z}\; = -\int_{-\infty}^{\infty}dq\Big[ | q \rangle \langle -q| \Big]  .
\end{equation}
The equality $\langle n|\Pi_{z}|n'  \rangle\;=\;\langle n |S_{z}|n'\rangle$ is easily verifiable. 
The natural vectorial operators that close an 
$su(2)$ algebra with $\Pi_{z}$ are
\begin{equation}
\Pi_{x}\;=\;\int_{0}^{\infty}dq\Big[ |{\cal E}\rangle\langle {\cal O}|\;+\;
|{\cal O}\rangle\langle{\cal E}|\Big] ,
\end{equation}
\begin{equation}
\Pi_{y}\;=\;i \int_{0}^{\infty}dq\Big[ |{\cal E}\rangle\langle {\cal O}|\;-\;
|{\cal O}\rangle\langle{\cal E}|\Big].
\end{equation}
We note that $\Pi_{x}$ is diagonal in $q$, i.e.,
\begin{equation}
\Pi_{x}\;\;=\;\;\int_{0}^{\infty}dq\Big[ |q  \rangle \langle q|\;-\;|-q  \rangle \langle  -q|\Big] 
= {\rm sgn} (\hat{q})
\end{equation}
is the spectral representation of $\Pi_{x}$.
Its representative Wigner function is
\begin{equation}
W_{\Pi_{x}}(q,p)\;=\;{\rm sgn}\;q ,
\end{equation}
i.e., it gives the eigenvalues ($\pm 1$) of the operator and hence is a proper (nondispersive) DV,
just as in the discussion of Eq. (\ref{A=f(q)}), case (b), of Sec. \ref{HVWT}.
In this case, with $\Pi_{i}$,
much like in the previous case (with the $S_{i},\;i\;=\;x,y,z$) it is easy to
get BIQV by selecting the appropriate orientational parameters. For convenience,
 while retaining complete generality, we consider the choice of the
orientational parameters by choosing the times (for both channels) of the
evolution of $\Pi_{x}^{1}(t_{1}), \Pi_{x}^{2}(t_{2})$ under the Hamiltonian
$H\;=\;\Pi_{z}$.
(We note that Bell \cite{bell2} considered the same
case with  $\zeta\;\rightarrow\;\infty$, i.e. the EPR state, but with the free Hamiltonian, $H\;=\;p^{2}/2$). 

Direct calculations show that by appropriate choice of the times ($t_{1},t_{1}'$ and
$t_{2},t_{2}'$) we get, for our case,
\begin{equation}
\langle{\hat B}\rangle\;=\;2\sqrt2{\bar F}(2\zeta);\;\; {\bar F}
\;=\;{2\over \pi}\arctan(\sinh2\zeta).
\label{<B>2nd level}
\end{equation}
Thus we see that in this case, where seemingly the quantal description may be given a LHV
underpinnng, we get a BIQV which, we are told, is an impossibility.
However, the present Bell operator involves not only the ``proper'' DV,
$\Pi_{x}$, but also $\Pi_{y}$ which evolves via our Hamiltonian, $H\;=\;\Pi_{z}$:
the latter, i.e., $\Pi_{y}$, is {\it not} a proper DV.
In fact, its Wigner representative is given by
\begin{equation}
W_{\Pi _y}(q,p) = -\delta (q)\;{\cal P}\frac 1p  ,
\end{equation}
where ${\cal P}$ stands for the ``principal value".
Thus, once again, no LHV underpinning for the correlation involved in Eq. (\ref{<B>2nd level})
is possible afterall
(the boundedness condition for the Wigner representatives
is violated as well).

We may attempt to consider the problem in a Schr\"odinger-like
manner by applying the time evolution operator to the state
$|\zeta\rangle$: this, however, leads to a new state, $|\zeta' \rangle$, whose Wigner
representative function is no longer non-negative over all phase space. This can be
proven most readily by considering an alternative expression for the state $|\zeta\rangle$
obtained in \cite{gour}, i.e.,
\begin{equation}
|\zeta \rangle
=\int_{o}^{\infty}\int_{0}^{\infty}\;dqdq' \Big[ (g_{+}\;+\;g_{-})|{\cal EE'}\rangle\;+\;(g_{+}\;-\;
g_{-})|{\cal OO'}\rangle\Big],
\label{tmss 1}
\end{equation}
where
\begin{eqnarray}
g_{\pm}(q,q,;\zeta)
&=&\langle qq'|S(\pm \zeta)|00\rangle
\nonumber \\
&=&{1 \over \sqrt \pi}\exp\Big\{-{1 \over 2}
\left[q^{2}\;+\;q'^{2}\;\mp2qq'\tanh(2\zeta)\right]
\nonumber \\
&&\hspace{1cm}\cdot \cosh(2\zeta)\Big\}.
\end{eqnarray}
Using this expression for $|\zeta\rangle$, we have directly
\begin{equation}
{\rm e}^{-i\gamma \Pi_{z}}|\zeta\rangle\;=\;|\zeta' \rangle\;=\;\cos\gamma|\zeta \rangle\;+\;\sin\gamma
|-\zeta \rangle,
\end{equation}
and the Wigner function representative of $|\zeta' \rangle$ is no longer non-negative \cite{schleich}.

\section{Bilinear Hamiltonians}
\label{bilinear}

Level 3 of our EPR-EPW problem is the study of cases wherein : (1) The states are having
non-negative Wigner representatives which, at some limit, reduce to the EPR state - our
$|\zeta\rangle$ is such a state. (2) A DV (= observable) that is nondispersive (=proper), i.e.,
such that the Wigner representative of its quantal version gives its eigenvalues in terms of
our LHV: $p,q$ - our $\Pi_{x}$ is such a DV.
We inquire for possible BIQV when this DV
evolves via Hamiltonians which leave the Wigner representative of the state under study
non-negative.
Alternatively, we inquire for BIQV when our
DV's evolve by Hamiltonians which allow the initially proper DV to remain as such.
In the next paragraphs
we study the relationship between these two alternatives.

Since the only
non-negative Wigner functions are gaussians \cite{schleich},
and as gaussians remain gaussians under linear transformations, single-channel Hamiltonians
that leave the Wigner function non-negative are bilinear ones.
We will consider now two such Hamiltonians:
\begin{subequations}
\begin{eqnarray}
H_{0}(i) &=& {1 \over 2}\left(\hat{p}^{2}_{i}\;+\omega _i ^2 \hat{q}^{2}_{i}\right)
\label{H ho}
\\
H_{f}(i)&=&{1 \over 2}\hat{p}^{2}_{i}  ,
\label{H f}
\end{eqnarray}
\end{subequations}
where the subscript $i=1,2$ denotes the channel.
For simplicity we shall consider, in $H_0$, the frequency $\omega_{i}=1$ for both channels.
The second Hamiltonian is the one considered by Bell \cite{bell2,lars}.

We consider the harmonic oscillator Hamiltonian $H_{0}$ first.
Evolution of the state $|\zeta \rangle$, Eq. (\ref{tmss}), under $H_0$,
during a time $t_1$ for channel 1 and $t_2$ for channel 2, gives:
\begin{equation}
\left|\zeta(t_{1},t_{2})\right\rangle \;=\;|\zeta' \rangle\;=\;\exp^{-\zeta
(a^{\dagger}b^{\dagger}e^{-i\theta}\;-\;abe^{i\theta})}|00\rangle,
\label{zeta(t)}
\end{equation}
where $\theta =t_{1}+t_{2}$.
The corresponding Wigner function
can be otained either directly from the state (\ref{zeta(t)}), or from Eq. (\ref{W tmss}), applying Eq. (\ref{W(rho') vs W(rho)}) with
$a=\cos t_i$, $b=\sin t_i$, $c=-\sin t_i$ and $d=\cos t_i$, with the result
\begin{eqnarray}
W_{\zeta(\theta)}\;
&=&\;{\frac{1 }{\pi^{2}}}\exp\Big\{-\cosh(2\zeta)\left(q_{1}^{2}+q_{2}^{2}+p_{1}^{2}+p_{2}^{2}\right)
\nonumber \\
&&\hspace{.8cm}-2\sinh(2\zeta)[(q_{1}q_{2}-p_{1}p_{2})\cos\theta
\nonumber \\
&&\hspace{1cm}-(q_{1}p_{2}+q_{2}p_{1})\sin\theta ]\Big\}.
\label{W(t) ho}
\end{eqnarray}
Direct evaluation of
\begin{equation}
E(t_{1},t_{2})=\int_{-\infty}^{\infty}dqdpW_{\zeta(\theta)}(q,p)\Pi_{x}^{1}\Pi_{x}^{2}
\label{E}
\end{equation}
($dqdp\;=\;dq_{1}dq_{2}dp_{1}dp_{2}$) gives (see App. \ref{E(t1t2)})
\begin{equation}
 E(t_{1},t_{2})\;=\;{\chi \over \pi},\;\;\;\;
\cos\chi\;=\;\tanh(2\zeta) \cos\theta.
\end{equation}
We have used the notation of \cite{bell2}
\begin{equation}
E(t_{1},t_{2}) \;=\;P_{++}(\theta)\;+\;P_{--}(\theta)\;-\;P_{-+}(\theta)\;
-\;P_{+-}(\theta).
\end{equation}
The first subscript refers to the eigenvalue (i.e., $\pm 1$) of $\Pi_{x}^{1}$
and the second subscript to that of the second channel    $\Pi_{x}^{2}$:
i.e., $P_{++}$ is the integral of $W_{\zeta(\theta)}(q,p)$ (see Eq. (\ref{E})) over the region
$q_1>0$, $q_2>0$, etc.
The alternative view, i.e., allowing $\Pi _x ^i$ to evolve in time, while keeping  $W_{\zeta }$ fixed, is readily done
(see App. \ref{Wigner Pix})
by noting that $\Pi _x ^i (t_i)={\rm sgn}(\hat{q}_i \cos t_i + \hat{p}_i \sin t_i)$
and computing the resulting integral for $E(t_{1},t_{2})$
\begin{equation}
E(t_{1},t_{2})=\int_{-\infty}^{\infty}dqdpW_{\zeta }(q,p)\Pi_{x}^{1}(t_1)\Pi_{x}^{2}(t_2)
\end{equation}
for this case upon the change of variables: 
$\bar{q}_i=q_i \cos t_i +p_i \sin t_i$ and
$\bar{p}_i=-q_i \sin t_i +p_i \cos t_i$. We obviously obtain the same answer at the end.
Perhaps more elegantly, one can find the Wigner representative of the time evolution of $\Pi _x ^i$ applying
the general result (\ref{W(f) vs W(f') 1}) of Sec. \ref{HVWT}, with 
$a=\cos t_i$, $b=\sin t_i$, $c=-\sin t_i$, and $d=\cos t_i$.

It is easily shown (cf.\cite{bell2}) that, in case the time dependence occurs only in the combination
$t_1 + t_2$ (which is the case in the present situation (Eq. (\ref{zeta(t)})),
the CHSH inequality \cite{shimony1} implies the following inequality
\begin{equation}
3P_{+-}(\theta)\;-\;P_{+-}(3\theta)\;\geq\;0.
\end{equation}
In the
$\zeta \rightarrow \infty$
limit, i.e., when the state
$\left|\zeta \right\rangle$
is  maximally entangled and approaches the EPR state, 
$\tanh(2\zeta)\;\rightarrow\;1$.
In this limit $\chi \rightarrow \cos^{-1}(\cos\theta)\;=\;\theta$ (cf. App. A) and
$P_{+-}(\theta)\;=\;{1 \over 2\pi}\theta$; thus the inequality is saturated.
It can be shown that for finite $\zeta$ the inequality is always satisfied.
Bell suggested that correlations of
observables of the type of $\Pi _x^{1,2}$ (cf. Eq. (\ref{E})) for the EPR state and
evolving under the free Hamiltonian would not allow for BIQV; we observe
that this indeed occurs for the harmonic oscillator Hamiltonian used here.

However, his reasoning perhaps was somewhat misleading:
the reason is that it is not only the nonnegativity of the relevant
Wigner function that matters, but also the type of evolution induced
in the observables
by the Hamiltonian in question.
The fulfillment of the CHSH inequality in the present case, in which the evolution is induced by the harmonic oscillator Hamiltonian,
is consistent with the discussion given in Sec. \ref{HVWT}, below Eq. (\ref{W(f) vs W(f') 1}).
It is apt to notice that the free Hamiltonian is not analogous to rotation of the spins in the Bohm
EPR version. The latter involves what was termed \cite{gour} orientational
variation, which leads (depending on the preferred viewpoint) either to
nonproper (dispersive) DV's even when ones starts with a proper DV, or,
alternatively, to a {\it non} nonnegative Wigner function. 
In either case, BIQV's do not contradict Bell's theorem.

We now consider briefly the evolution due to the free Hamiltonian of Eq. (\ref{H f}).
Again, we study the evolution of the state $|\zeta \rangle$, Eq. (\ref{tmss}), under  $H_f$,
during a time $t_1$ for channel 1 and $t_2$ for channel 2.
The corresponding Wigner function can be otained from Eq. (\ref{W tmss}), applying Eq. (\ref{W(rho') vs W(rho)}) with
$a=1$, $b=t_i$, $c=0$ and $d=1$,
with the result
\begin{widetext}
\begin{eqnarray}
W_{\zeta(t_1, t_2)}
&=&{\frac{1 }{\pi^{2}}}
\exp\bigg\{-\cosh(2\zeta)\left[(q_1-t_1p_1)^{2}+(q_2-t_2p_2)^{2}+p_{1}^{2}+p_{2}^{2}\right]
\nonumber \\
&&\hspace{1cm}- 2\sinh(2\zeta)\Big[(q_1-t_1p_1)(q_2-t_2p_2)-p_{1}p_{2} \Big]\bigg\}.
\label{W(t) f}
\end{eqnarray}
\end{widetext}
With the same definitions as above, we find
\begin{subequations}
\begin{eqnarray}
E(t_1, t_2) =\frac{2}{\pi } \arcsin \left[\alpha (t_1, t_2)\tanh 2\zeta \right]  ,
\label{E f}
\\
\alpha (t_1, t_2) = \frac{1-t_1t_2}{\sqrt{(1+t_1^2)(1+t_2^2)}}  .
\label{alpha}
\end{eqnarray}
\end{subequations}
Alternatively, just as with the previous Hamiltonian $H_0$, one can find the Wigner representative of the time evolution of
$\Pi _x ^i$ applying
the general result (\ref{W(f) vs W(f') 1}) of Sec. \ref{HVWT}, with $a=1$, $b= t_i$, $c=0$, and $d=1$.

We wish to analyze whether this problem abides by the CHSH inequality, i.e., whether the inequality
\begin{equation}
\left|
E(t_1, t_2) + E(t_1, t'_2) +E(t'_1, t_2) - E(t'_1, t'_2)
\right| \leq 2 ,
\label{CHSH for Hf}
\end{equation}
or
\begin{eqnarray}
&&\Big|
\arcsin \left[\alpha (t_1, t_2)\tanh2\zeta \right]  + \arcsin \left[\alpha (t_1, t'_2)\tanh2\zeta \right]
\nonumber \\
&&\hspace{.8cm}+\arcsin \left[\alpha (t'_1, t_2)\tanh2\zeta \right]
\nonumber \\
&&\hspace{1cm} - \arcsin \left[\alpha (t'_1, t'_2)\tanh2\zeta \right]
\Big|
\leq \pi ,
\label{CHSH for Hf 1}
\end{eqnarray}
is satisfied. For instance, taking
\begin{equation}
\begin{array}{cccc}
t_1=0, & t'_1=T, & t_2=0, & t'_2 =T ,
\end{array}
\end{equation}
and subsequently taking the limit $T \rightarrow \infty$, the left-hand side of (\ref{CHSH for Hf}b) takes the value $\pi $,
i.e., the inequality is saturated.

The fulfillment of the CHSH inequality in the present case, in which the evolution is induced by the free Hamiltonian,
is, once again, consistent with the discussion given in Sec. \ref{HVWT},
below Eq. (\ref{W(f) vs W(f') 1}).

\section{Conclusions and Remarks}
\label{concl}

In this study we took the Clauser, Horne, Shimony and Holt \cite{shimony1} inequality as the
representative of the so called Bell's inequalities.
Indeed this inequality is the often analyzed and experimentally
tested one and is the one used by Bell himself in his study of the subject of this work:
the relation of the non-negative Wigner function of the Einstein, Podolsky and Rosen state to possible
Bell's inequality violations.
Our results are mundane: no violation is possible when such is not to be allowed by Bell's inequality.
We subjected the reader to a lengthy derivation and explanation of what we
considered points worthy of clarification.
These were the delineation of what is meant by proper and
improper dynamical variables in the context of the Wigner function as a probability distribution
in phase space, the canonical variables of the latter playing the role of the local hidden variables,
and showed that a proper observable (= dynamical variable) is non-dispersive.
Thus  only proper
dynamical variables can be considered as accountable for by a local hidden variable theory with the
phase space variables ($q,p$) being the local hidden variables. A proper dynamical variable is
one whose Wigner function representative gives the eigenvalues of the corresponding quantal dynamical
variable which the local hidden variable theory aims at underpinning.

Now, although the word ``local" was repeated several times, locality as such was not an issue
in the present discussion: 
Bell's locality condition is automatically fulfilled as the Wigner function of any dynamical variables that
depend on distinct phase space coordinates factorizes. Thus our discussion underscores a tacit assumption in the derivation of the Bell inequality we consider: viz the dynamical variables
must all have a definite value even though they are not or even cannot be measured simultaneously.
This point was noted in the past \cite{wigajp,sakurai,willy,stapp,laloe}.
In point of fact, two often quoted examples for underpinning non-commuting dynamical variables with
LHV's 
--Bell's \cite{bell1} and Wigner's \cite{wigajp}-- are manifestly so, although these examples are, perhaps,
somewhat artificial.
In the present work
--which in its essence follows Bell's suggestion \cite{bell2}--
we outlined a canonical theory which automatically abides by the locality requirement
(the phase space variables are local), and BIQ is abided by in cases where the DV's are proper ones, even when they are non-commuting.

Our main conclusion is that the validity of Bell's inequality
that we have considered hinges on the assumption of having definite values for all the
dynamical variables --thus endowing them with physical reality-- and not the issue of locality.
Of course one might ponder what would one mean
by a local hidden variable theory without a definite value for all the 
dynamical variables; however this is a separate issue.

\begin{acknowledgments}
One of us (MR) expresses its gratitude to the Elena Aizen de Moshinsky Chair, through whose financial
support his visit to UNAM (where a large part of this article was produced) was made possible.
\end{acknowledgments}

\appendix

\section{Evaluation of $E(t_{1}, t_{2}) $ for the Harmonic Hamiltonian}
\label{E(t1t2)}

We first evaluate $P_{-+}(t_{1},t_{2}), $ cf. Eq. (61). The integral, Eq. (59), {\it after} the integration
over the $p$'s and letting $q_{1}\;\rightarrow\;-q_{1}$, is
\begin{eqnarray}
&&P_{-+}(t_{1},t_{2})
={1 \over \pi \cosh(2\zeta)\sqrt{(1-\tanh^{2}(2\zeta) cos^{2}\theta)}}
\nonumber \\
&&\hspace{1cm}\cdot\int_{0}^{\infty}dq_{1}dq_{2}\exp\Big[-\cosh(2\zeta)\Gamma(\theta,\zeta)
\nonumber \\
&&\hspace{1.5cm}\cdot\left(q_{1}^{2}+q_{2}^{2}-
2q_{1}q_{2}\tanh(2\zeta)cos\theta\right)\Big].
\nonumber \\
\end{eqnarray}
Here  $\theta = (t_{1}\;+\;t_{2})$ and  $\Gamma(\theta,\zeta)\;=\;(1-\tanh(2\zeta))/(1-\tanh(2\zeta)cos^{2}
\theta)$. This integral is evaluated directly to give
\begin{eqnarray*}
P_{-+}(t_{1},t_{2}) &=&{1 \over 2\pi}\bigg[
{\pi \over 2}
\nonumber \\
&-& \arctan\left(
{\tanh(2\zeta)\cos\theta \over \sqrt{(1-\tanh^{2}(2\zeta)cos^{2}\theta)}}
\right)
\bigg].
\end{eqnarray*}
Similar calculation gives for
\begin{eqnarray*}
P_{++}(t_{1},t_{2}) &=&{1 \over 2\pi}\bigg[
{\pi \over 2}
\nonumber \\
&+&
\arctan\left(
{\tanh(2\zeta)\cos\theta \over \sqrt{(1-\tanh^{2}(2\zeta)cos^{2}\theta)}}
\right)
\bigg].
\end{eqnarray*}
The equality $P_{++}(t,t')\;=\;P_{--}(t,t')$ and  $P_{+-}(t,t')\;=\;P_{-+}(t,t')$  is easily verifiable,
hence we have for
\begin{equation}
E(t_{1},t_{2})\;=\; 2P_{++}(\theta)\;-\;2P_{-+}(\theta)\;=\;{\chi \over \pi},
\end{equation}
with $\tanh(2\zeta)\cos \theta\;\equiv\;\cos\chi,\;\theta\;=\;t_{1}+t_{2}$.

\section{The Wigner Function of $\Pi_{x}(t)$ for $H\;=\;H_{0}$}
\label{Wigner Pix}

The Wigner function for $\Pi_{x}(t)$ is given by
\begin{eqnarray}
&&W_{\Pi_{x}(t)}(x,p)
={1 \over 2\pi}\int_{-\infty}^{\infty}dy\int_{0}^{\infty}dqe^{-ipy}
\nonumber \\
&&\hspace{1cm}\cdot \left\langle x+y/2 \right| e^{iH_{0}t}
\nonumber \\
&&
\hspace{1.5cm}\cdot\Big[\left|q\right\rangle  \left\langle q\right|
-\left| -q \right\rangle \left\langle -q \right|  \Big] e^{-iH_{0}t} \left| x-y/2 \right\rangle .
\nonumber \\
\end{eqnarray}
Inserting the harmonic oscillator propagators \cite{schulman} and performing
the $y$ integration gives
${\rm sgn} (x\cos t + p\sin t)$.


\begin{thebibliography}{999}

\bibitem{bell2} J. S. Bell, {\it Speakable and Unspeakable in Quantum Mechanics}  (Cambridge
 U. press, Cambridge, UK, 1987), p. 196.
\bibitem{bell1}J. S. Bell, {\it Speakable and Unspeakable in Quantum Mechanics}  (Cambridge
U. press, Cambridge, England, 1987), p. 14.
\bibitem{rosen} A. Einstein, B. Podolsky and N. Rosen, Phys. Rev. {\bf 47}, 777 (1935).
\bibitem{wigner} E. P. Wigner, Phys. Rev. {\bf 40}, 749 (1932).
\bibitem{ulf} U. Leonhardt, {\it Measuring the Quantum State of Light}, (Cambridge U. Press,
Cambridge, UK, 1997).
.
\bibitem{shimony2} J. F. Clauser and A. Shimony, Rep. Prog. Phys. {\bf 41}, 1881 (1978).
\bibitem{faqir} F. C. Khanna, A. Mann, M. Revzen and S. Roy, Phys. Lett. A
{\bf 294}, 1 (2002).
\bibitem{potasek} B. Yurke and M. Potasek, Phys. Rev
. A {\bf 36}, 3464 (1987).
\bibitem{wodkiewicz}K. Banaszek and K. Wodkiewicz, Phys. Rev. A {\bf 58};
4345 (1998), Phys. Rev. Lett. {\bf 82}, 2009 (1999).
\bibitem{70} Recent Developments in Quantum Physics, Feb 1 - 2. 2004 (Conference in honor of Asher Peres' 70th birthday) Technion, Haifa, Israel. 
\bibitem{shimony1} J. F. Clauser, M. A. Horne, A. Shimony and R. A. Holt,
Phys. Rev. Lett. {\bf23}, 880 (1969).
\bibitem{sam} S. L. Braunstein, A. Mann and M. Revzen. Phys. Rev. Lett. {\bf 68}, 3259 (1992).
\bibitem{lars} L. M. Johansen, Phys. Lett. A, {\bf 236}, 173 (1997).
\bibitem{reid} M. D. Reid and  D. F. Walls , Phys. Rev. A {\bf 34}, 1260 (1986).
\bibitem{walls} D. F. Walls and G. J. Milburn, {\it Quantum Optics} (Spinger-Verlag, Berlin,1994).
\bibitem{grangier} P. Grangier, M. J. Potasek and B. Yurke, Phys. Rev. A, {\bf 38}, 3132 (1988).
\bibitem{mandel} A. Kuzmich, I. A. Walmsley and L. Mandel, Phys. Rev. Lett.
{\bf 85}, 1349 (2002).
\bibitem{ou1} Z. Y. Ou, S. F. Pereira and H. J. Kimble, Phys. Rev. Lett. {\bf 68}, 3663 (1992).
\bibitem{ou2} Z. Y. Ou, S. F. Pereira and H. J. Kimble, App. Phys. B {\bf 55}, 265 (1992).
\bibitem{santos} A. Casado, T. W. Marshall and E. Santos, J. Opt. Soc. Am.B,
 {\bf 15}, 1572 (1998).  
\bibitem{zeng} Zeng-Bing Chen, Jian-Wei Pan GuangHou and Yong-De Zhang, Phys. Rev. Lett.,
 {\bf 88}, 040406-1 (2002).
\bibitem{cirelson} B. S. Cirelson, Lett. Math. Phys. {\bf 4}, 93 (1980).
\bibitem{landau} L. J. Landau, Phys. Lett. A {\bf 120}, 54 (1987).
\bibitem{entangled} In this limit the probability of an even parity state equals that of
an odd parity state, cf. Eq. (\ref{tmss 1}) below. It is for such a state that maximal BIQV is attainable; see, e.g.,
N. Gisin, Phys. Lett. A {\bf 154}, 201 (1991).
\bibitem{gour} G. Gour, F. C. Khanna, A. Mann and M. Revzen, Phys. Lett. A {\bf  324}, 415 (2004).
\bibitem{agrawal} Examples of alternative domains were considered in
J. A. Vaccaro and D. T. Pegg, Phys. Rev. A {\bf 41}, 5156 (1990);
G. S. Agarwal, D. Home and W. Schleich, Phys. Lett. A  {\bf 170}, 359 (1992).
\bibitem{wigajp}E. P. Wigner, Am. Jour. Phys. {\bf 38}, 1005 (1970).
\bibitem{sakurai}J. J. Sakurai, {\it Modern Quantum Mechanics, Revised Edition} (Addison - Wesley, Reading, Mass., 1985), p. 227
\bibitem{willy} W. De Baere, A. Mann and M. Revzen, Found. Phys. {\bf 29}, 67 (1999).
\bibitem{milos} M. Revzen, M. Lokajicek and A. Mann, Quant. Semiclas. Opt, {\bf 9}, 501 (1997).
\bibitem{stapp} H. Stapp, Am. J. Phys. {\bf 72}, 30, (2004).
\bibitem{laloe} F. Lalo$\ddot e$, Am. J. Phys. {\bf 69}, 655 (2001).
\bibitem{schleich1}W. Schleich, {\it Quantum Optics in Phase Space}, Wiley-VCH,
 Berlin (2001), p. 89.
\bibitem{schleich} R. L. Hudson, Rep. Math. Phys. {\bf 6}, 249 (1974);
W. Schleich, {\it Quantum Optics in Phase Space}, Wiley-VCH,
 Berlin (2001), p. 74.

\bibitem{lars2}A nondispersive DV is not necessarily a proper DV. The latter
should be used as the requirement for hidden variable underpinning whenever 
distinction arises.
\bibitem{moshinsky}M. Moshinsky and C. Quesne, Jour. Math. Phys. {\bf 12}, 1772 (1971). J. G. Kr$\ddot u$ger and A. Poffyu, Physica A, {\bf 91}, 99 (1978).
\bibitem{schulman} L. S. Schulman {\it Techniques and Applications of Path Integrals}, (Wiley Interscience, New York, 1981).

\end{thebibliography}
\end{document}